\documentstyle[11pt,moriond,epsfig]{article}

\def\be{\begin{equation}}
\def\ee{\end{equation}}
\def\bea{\begin{eqnarray}}
\def\eea{\end{eqnarray}}

\begin{document}
\title{QUANTUM DECOHERENCE AND THE ``P(E)-THEORY''}

\author{D.S. GOLUBEV$^a$ and A.D. ZAIKIN$^{b,a}$ }

\address{$^a$ P.N.Lebedev Physics Institute, Leninskii pr. 53, 117924 Moscow, 
Russia\\
$^b$ Forschungszentrum Karlsruhe, Institut f\"ur Nanotechnologie,
76021 Karlsruhe, Germany}

\maketitle\abstracts{We point out a close physical and formal similarity 
between the
problems of electron tunneling in the effective environment and the
weak localization effects in the presence of interactions. In both cases
the results are expressed in terms of the ``energy probability distribution
function'' $P(E)$ which has a finite width even at $T=0$ due to interactions.}

\section{Introduction}
Recent experiments \cite{Webb} strongly indicate an intrinsic nature of
the low temperature saturation of the electron decoherence time $\tau_{\varphi}$
in disordered conductors \cite{AAK,AAK1}, thereby attracting a lot of 
attension to the
fundamental role of interactions in such systems. A theory of the above
phenomenon was proposed in \cite{GZ1,GZ2} where it was demonstrated that
electron-electron interactions in disordered systems are indeed responsible
for a nonzero electron decoherence rate down to $T=0$. It was argued 
\cite{GZ98} that this interaction-induced decoherence has the
same physical nature as in the case of a quantum particle with 
coordinate $q$ interacting with a bath of harmonic oscillators. 

The latter problem can be described
by the exactly solvable Caldeira-Leggett model \cite{CL2}. Within this model
one can easily observe that the off-diagonal elements of the particle
density matrix $\rho (q_1,q_2)$ decay and even at $T=0$
in equilibrium they are suppressed at a typical length scale 
$L_d \sim 1/\sqrt{\langle p^2 \rangle }$, where the expectation value
for the square of the particle momentum operator $\langle p^2 \rangle $
is determined by the interaction strength and remains nonzero \cite{CL2,HA}
even {\it in the true ground state} of the total system 
``particle+oscillators''. It is important to emphasize that the decoherence 
length $L_d$ characterizes the behavior of the particle $q$ and not
of the (obviously coherent) eigenmodes of the total system obtained by
an exact diagonalization of the initial Hamiltonian \cite{HA}. However, 
if the behavior of the particle
$q$ (and not of the system eigenmodes) is of interest, the reduced
density matrix $\rho (q_1,q_2)$ should be studied and the finite decoherence
length $L_d$ will explicitly enter into the expectation values for the
corresponding operators. 
  
Similarly, in order to evaluate the conductance of a disordered metal 
in the presence of electron-electron 
interactions it is convenient to choose the basis of {\it noninteracting} 
electrons simply because the current operator
can be easily expressed in this basis. Then, as in the previous example,
the reduced single electron density matrix in the presence of interactions
should be evaluated, and after that the expression for the
system conductance is readily established. This program was carried 
out in \cite{GZ2} where the interplay between interaction and weak localization 
effects was nonperturbatively studied by means of the real time path integral 
technique. 
 
The aim of this paper is to highlight a similarity between our theory 
\cite{GZ1,GZ2} and the so-called ``$P(E)$-theory'' \cite{Naz}
describing electron tunneling in mesoscopic tunnel junctions in the
presence of interaction with the effective electronic environment in
metallic electrodes. We will demonstrate that both results \cite{GZ1,GZ2}
and \cite{Naz} naturally follow from the same nonperturbative
procedure \cite{GZ2}.

\section{Microscopic Path Integral Analysis}

Our starting point is the standard Hamiltonian for the interacting
electrons in a disordered metal $\hat{H}_{el}=\hat{H}_0+\hat{H}_{int}$:
\begin{equation}
\hat{H}_0= \int dr\psi^+_{\sigma}(r)
\left[-\frac{\nabla^2}{2m}-\mu +U(r)\right]\psi_\sigma (r),
\label{H0}
\end{equation}
\begin{equation}
\hat{H}_{int}=\frac{1}{2}\int dr\int 
dr'\psi^+_{\sigma}(r)\psi^+_{\sigma'}(r')
\frac{e^2}{|r-r'|}\psi_{\sigma'}(r')\psi_{\sigma}(r).
\label{Hint}
\end{equation}
Here $U(r)$ includes both the random potential of impurities and the potential
of a tunnel barrier if the latter is present. Performing the 
Hubbard-Stratonovich transformation in the interaction term (\ref{Hint})
we reformulate the initial many-body problem in terms of a single electron
interacting with a two-component quantum field $V_{1,2}$. The exact electron
propagator $\hat G$ (which is $2\times 2$ matrix in the Keldysh space) is
expressed in terms of the propagator $\hat G_V$ for a single electron in the
field $V_{1,2}$ as follows 
\begin{equation}
{\hat G}=\frac{\int{\cal D}V_1{\cal D}V_2
\; \hat G_V\; e^{iS[V_1,V_2]}}
{\int{\cal D}V_1{\cal D}V_2\; e^{iS[V_1,V_2]}},
\label{GV0}
\end{equation}
where the dynamics of the field $V$ is determined by the effective action
\begin{equation}
iS[V_1,V_2]=2{\rm Tr}\ln\hat G_V^{-1}+ i\int\limits_0^t dt'\int
dr \frac{(\nabla V_1)^2-(\nabla V_2)^2}{8\pi}.
\label{S}
\end{equation}
Note, that eq. (\ref{S}) is obtained by integrating out the Grassman electron 
fields and, hence, it explicitly accounts for the Fermi statistics 
(see \cite{GZ2} for more details).

To proceed further we will evaluate the action (\ref{S}) outside the
tunnel junction within RPA (which amounts to expanding 
${\rm Tr}\ln\hat G_V^{-1}$ in $V$ up to terms $\sim V^2$) and find the
tunnel barrier contribution to $S$ by means of the procedure \cite{AES,SZ}.
Then we obtain 
\begin{eqnarray}
iS&=&
i\int\frac{d\omega d^dk}{(2\pi)^{d+1}}
V^-(-\omega,-k)\bigg[C(\omega,k)+
\frac{a^{3-d}k^2
\big(\epsilon(\omega,k)-1\big)}{4\pi}\bigg]V^+(\omega,k)-
\nonumber \\
&&
-\frac{a^{3-d}}{2}
\int\frac{d\omega d^3k}{(2\pi)^4}
V^-(-\omega,-k)\frac{k^2{\rm Im}\epsilon(\omega,k)}{4\pi}
\coth\left(\frac{\omega}{2T}\right)
V^-(\omega,k) +S_{AES}[\varphi_+,\varphi_-]. 
\label{RPA}
\end{eqnarray}
Here $a$ is the transverse size of 1d and 2d systems, 
$C(\omega ,k)$ is the effective capacitance,
\begin{equation}
\epsilon(\omega,k)=1+\frac{8\pi e^2N_0D}{-i\omega +Dk^2}
\label{eps}
\end{equation}
is the dielectric susceptibility, $N_0=mp_F/2\pi^2$ is the electron density
of states, $D=v_Fl/3$ and $S_{AES}$ describes tunneling through
the barrier and has the well known form \cite{AES,SZ}. The field $\varphi^{\pm}$
is related to the jump of the $V$-field across the barrier 
in a standard manner \cite{SZ} $\dot \varphi^{\pm}/2e=V_{+0}^{\pm}-V_{-0}^{\pm}$.
If the tunnel barrier is absent, the $V$-field is continuous, 
$\dot \varphi^{\pm}=0$, and $S_{AES}=0$.  

Let us evaluate the conductivity of the system in both cases, i.e.
with and without the tunnel barrier. In order to do that we start from the
standard quantum mechanical expression for the current operator and after
some formal manipulations (see \cite{GZ2} for more details) express 
the system conductivity in terms of the path integral with the effective
action (\ref{RPA}), see eq. (49) of \cite{GZ2}. We will proceed in parallel
in order to illustrate the analogy between the two cases at each step of
the calculation. 

{\it Tunnel junction}. If the tunnel junction is present in the system we 
will be interested only in its contribution to the conductance. Expanding
the path integral in $S_{AES}$ (see e.g. \cite{Naz,SZ} for details) we
obtain the expression for the system conductance 
\begin{equation}
G(V)=\frac{1}{R_T^{(0)}}-\frac{2}{\pi R_T^{(0)}}\int\limits_0^{+\infty}dt\;\;
t\left(\frac{\pi T}{\sinh\pi Tt}\right)^2{\rm Im}\left[P(t)\right]\cos(eV_xt)
\label{GV}
\end{equation}
Here $R_T^{(0)}$ is the ``classical'' resistance of the tunnel junction, $V_x$ is
the applied voltage and the function $P(t)$ describes energy smearing for 
tunneling electrons due to interaction with the fluctuating
field $V$ produced by other electrons. This function has the form
\begin{equation}
P(t)=\left\langle e^{i\int_0^t dt'\; \left(f^-V^++f^+V^-\right)} 
\right\rangle_{V^+,V^-}, \;\; f^-=e, \;\; f^+=e/2.
\label{PTJ}
\end{equation}
The expression in the exponent (\ref{PTJ}) has a simple physical
meaning: It describes the phase acquired by a tunneling electron in the
field $V$. The correlator for this field at the junction is obtained from
the action (\ref{RPA}) (with $S_{AES}=0$) in the limit $k \to 0$. Assuming that
the capacitance is dominated by that of a tunnel junction $C$ and 
defining $V^{\pm}_{\omega}=i\omega \varphi_{\omega}/2e$ one readily
finds
\begin{equation}
\langle|V^+_\omega|^2\rangle=\frac{\omega\coth\left(\frac{\omega}{2T}\right)}
{\frac{(\omega C)^2}{G_s}+G_s}, \;\;\;\;\; G_s=\sigma ^{(0)}a^2/L,
\label{V+V+1}
\end{equation}
where $\sigma^{(0)}=2e^2N_0D$ and $L$ is the length of a disordered 
conductor shunting the tunnel junction. 

{\it Disordered Conductor}. Now let us eliminate a tunnel junction and
evaluate the conductivity of a disordered metal in the presence
of weak localization and interaction effects. Again making use of eq. (49)
of \cite{GZ2} and disregarding the so-called interaction correction \cite{AAK1}
we obtain 
\begin{equation}
\sigma=\sigma^{(0)}-\frac{2e^2D}{\pi}\int\limits_{\tau_e}^{\infty}
dt\;\; \frac{\langle P(t)\rangle_{\rm diff}}
{\left(4\pi Dt\right)^{d/2}a^{3-d}}
\label{sigma}
\end{equation}
where (see \cite{GZ2} for details)
\begin{equation}
P(t)=\left\langle e^{i\int_{0}^tdt'\int dr
\big(f^-V^+ + f^+V^-\big)}\right\rangle_{V^+,V^-},
\label{PDC}
\end{equation}
\begin{eqnarray}
f^-(t',r)&=&e\delta (r-r_1(t'))-e\delta (r-r_2(t')),
\nonumber \\
f^+(t',r)&=&\frac{1}{2}
\bigg(e\big[1-2n(\xi_1(t'))\big]\delta
(r-r_1(t'))+e\big[1-2n(\xi_2(t'))\big]\delta (r-r_2(t'))\bigg).
\label{fpm}
\end{eqnarray}
Here $r_{1,2}(t)$ are electron trajectories with energies $\xi_{1,2}$, 
and $n(\xi )$ is the Fermi function.
As in the case of a tunnel junction the (Fourier transformed) function $P(t)$
describes energy smearing for an electron propagating in a disordered
conductor and interacting with the fluctuating field $V(\omega ,k)$ produced
by other electrons. The correlators for this field are again defined by the
action (\ref{RPA}). E.g. for the correlator $\langle V^+V^+ \rangle$ in the
most interesting limit $C(\omega ,k)D \ll \sigma_d$ we find 
\begin{equation}
\langle|V^+_{k,\omega}|^2\rangle=a^{3-d}
\frac{\omega\coth\left(\frac{\omega}{2T}\right)}
{\frac{(\omega C(\omega,k))^2}{\sigma_d k^2}+
\sigma_d k^2},   \;\;\;\; \sigma_d=\sigma^{(0)}a^{3-d}.
\label{V+V+2}
\end{equation}
We observe an obvious similarity between the results (\ref{GV}) and 
(\ref{sigma}). In both cases the quantum correction to the conductance
is expressed in terms of the function $P(t)$ which has essentially
the same form (cf. eqs. (\ref{PTJ}) and (\ref{PDC})) and the same
physical meaning: it accounts for the phase accumulation of the electron
propagating in the fluctuating field $V$. The electron trajectories differ
in these two cases (they are confined to the junction area in the case of
tunneling electrons and they are extended in space for electrons propagating
in a disordered conductor), however this difference is purely quantitative
and is completely unimportant for our comparison. An additional (and also
unimportant) difference is that averaging over diffusive trajectories
is carried out in eq. (\ref{sigma}) while no such averaging is needed in eq.
(\ref{GV}). Finally, we observe that the correlators
for the fluctuating field (\ref{V+V+1}) and (\ref{V+V+2}) also have essentially
the same form. This equivalence is by no means surprizing, since both
expressions follow from the same fluctuation-dissipation theorem for
disordered conductors. The same equivalence exists between the correlators
$\langle V^+V^- \rangle$ which we do not present here.

Let us now evaluate the expressions (\ref{GV}) and (\ref{sigma}) by averaging
over the fluctuating field $V$ in (\ref{PTJ}) and (\ref{PDC}). This averaging
is Gaussian in both cases and, hence, can be carried out exactly.    

{\it Tunnel junction}. Integrating out the fluctuating field $V$ in (\ref{PTJ}) 
we arrive at the well known result \cite{Naz}
$$
\ln P(t)= \frac2g\int_0^{\infty}\frac{d\omega}{\omega }\frac1{1+(\omega /\omega_c)^2}
\left(\coth \left(\frac{\omega}{2T}\right)[\cos (\omega t)-1]-
i \sin (\omega t)\right) 
$$
\begin{equation}
\simeq -\frac{2\pi}{g}Tt- \frac2g\ln \frac{1-e^{-2\pi Tt}}{2\pi (T/\omega_c)} 
-\frac{i\pi}{g},\;\;\;\;\;\;\; g=2\pi G_s/e^2.
\label{P(t)} 
\end{equation}
In the limit max$(eV_x,T) \ll \omega_c=G_s/C$ this equation
together with (\ref{GV}) yields \cite{PZ,SZ,Naz}
\begin{equation}
GR_T^{(0)} \propto {\rm max}[(T/\omega_c)^{2/g},(eV_x/\omega_c)^{2/g}]. 
\label{zba}
\end{equation}

{\it Disordered Conductor}. Integrating out the fluctuating field $V$ in 
(\ref{PDC}), substituting the time reversed paths $r_1(t)$ and $r_2(t)$
into (\ref{fpm}) and averaging the result over diffusive trajectories
we find (see also \cite{GZ1,GZ2}) $\langle P(t)\rangle_{\rm diff}\simeq 
\exp (-f_d(t))$, where \cite{GZ99}
\begin{eqnarray}
f_d(t)&=&
\frac{4e^2D^{1-d/2}}{\sigma_d (2\pi)^d}
\left(\int\frac{d^dx}{1+x^4}\right)
\int\frac{d\omega\; d\omega'}{(2\pi)^2}
\left[\frac{|\omega'|^{d/2-2}(\omega-\omega')\coth\frac{\omega-\omega'}{2T}
}{\omega^2}
\right.
\nonumber\\
&&
\left.
+
\frac{|\omega'|^{d/2-2}\omega\coth\frac{\omega}{2T}+
|\omega|^{d/2-2}\omega'\coth\frac{\omega'}{2T}}{\omega^{\prime 2}-\omega^2}
\right](1-\cos\omega t).
\label{fott}
\end{eqnarray}
Let us consider the case of quasi-1d conductors. From (\ref{fott}) we obtain 
\cite{GZ99}
\begin{equation}
f_1(t)=\frac{e^2}{\pi\sigma_1}\sqrt{\frac{2D}{\tau_e}}t+
\frac{2e^2}{\pi\sigma_1}\sqrt{\frac{Dt}{\pi}}\left(
\ln\frac{2\pi t}{\tau_e}-6\right), \;\;\; \pi Tt\ll 1,
\label{fquantum}
\end{equation}
\begin{eqnarray}
f_1(t)&=&\frac{2e^2}{\pi\sigma_1}\sqrt{\frac{D}{\pi}}\left\{
\sqrt{\frac{\pi}{2\tau_e}}t+\frac{2\pi}{3}Tt^{3/2}+\frac{\pi\zeta(1/2)}
{\sqrt{2}}t\sqrt{T}
-\frac{3\zeta(3/2)}{4\sqrt{2}}\frac{1}{\sqrt{T}}
\right.
\nonumber\\
&&
\left.
+
\sqrt{t}\ln\left(\frac{1}{4T\tau_e}\right)
+{\cal O}\left(\frac{1}{T\sqrt{t}}\right)+
2\pi Tt^{3/2}e^{-2\pi Tt}+
{\cal O}(\sqrt{t}e^{-2\pi Tt})\right\},\;\;\; \pi Tt\gg 1,
\label{fthermal}
\end{eqnarray}
where $\tau_e=v_F/l$ and $\zeta (x)$ is the dzeta-function. Making use of (\ref{sigma}), (\ref{fquantum}) and (\ref{fthermal}) we arrive 
at the standard form of the weak localization correction in 1d: 
$\sigma - \sigma^{(0)} \simeq (e^2/\pi a^2) \sqrt{D\tau_{\varphi}}$,
where \cite{GZ1,GZ2,GZ98} $1/\tau_{\varphi} = 
(e^2/\pi \sigma_1)\sqrt{2D/\tau_e}$ for $T \ll 1/\sqrt{\tau_e\tau_{\varphi}}$ 
and \cite{AAK,AAK1} $1/\tau_{\varphi} \sim (e^2D^{1/2}T/\sigma_1)^{2/3}$ 
for $T \gg  1/\sqrt{\tau_e\tau_{\varphi}}$. 

\section{Discussion}

As follows from the above analysis, in both physical situations considered here
the electron-electron interaction plays essentially the same role: Due to the 
energy
exchange between propagating/tunneling electrons and an {\it intrinsic} 
fluctuating electromagnetic field produced by other electrons the electron
energy $E$ is smeared even at $T=0$. This smearing is described by the function
$P(t)$ which Fourier transform $P_E$ plays the role of the energy probability
distribution. For the electrons tunneling through the barrier at $T=0$ and 
$E\geq 0$ one has \cite{Naz}
\begin{equation}
P_E = \frac1{2\pi}\int_{-\infty}^{+\infty}dt e^{iEt}P(t) \sim E^{-1+2/g},
\label{PE1}
\end{equation}
while for quasi-1d disordered conductors the analogous function evaluated
on the time reversed diffusive paths (again at $T=0$ and $E \geq 0$) reads
\begin{equation}
P_E=\frac1{2\pi}\int_{-\infty}^{+\infty}dt e^{iEt}\langle P(t)\rangle_{\rm diff}
\sim \frac{\tau_\varphi}{1+E^2\tau_\varphi^2}. 
\label{PE2}
\end{equation}
Note that without interaction the function
$P_E$ reduces to $P_E = \delta (E)$
in both cases (\ref{PE1}) and (\ref{PE2}). {\it Electron-electron
interaction is responsible for smearing of the energy distribution P(E) and
for a nonzero decoherence rate $1/\tau_{\varphi}$ at T=0}.

It is also important to emphasize that in both cases a nonperturbative
analysis of exactly the same type was used in order to account for
the interaction effects. These effects can be easily mistreated or even
completely missed by insufficient approximations. A clear illustration
of this point is provided by the ``$P(E)$-theory'' results (\ref{P(t)}), 
(\ref{zba}). E.g. it is obvious that a simple perturbative expansion
in $1/g$ in (\ref{P(t)}), 
(\ref{zba}) (essentially equivalent to the short 
time expansion) is insufficient and would lead to divergent results at
small $T$ and $V_x$ because of the nonanaliticity in $1/g$. On the other 
hand, the long time expansion (essentially equivalent to the golden-rule-type 
approximation)
would also yield an incorrect result at low $T$: E.g. smearing (\ref{PE1})
at $T \to 0$ would not be captured if one would expand the result 
(\ref{P(t)}) to any finite order in $1/Tt$. This procedure would yield an 
incorrect conclusion that at low $T$ the function $P_E$ has an effective 
width $\propto T$ while the remaining term in (\ref{P(t)}) would be a 
$t$-independent constant $\sim \ln (\omega_C/T)$. A truly
nonperturbative procedure is needed to obtain the correct results
(\ref{P(t)}), (\ref{zba}).

A similar conclusion can be drawn for the weak localization correction
in the presence of interactions. In this case the full expression for the
$P$-function has the form \cite{GZ99} $\langle P(t)\rangle_{\rm diff}=A_d(t)
\exp (-f_d(t))$, where $A_d(t)$ also depends on the 
interaction. On the relevant time scales $t \sim \tau_{\varphi}$ the 
pre-exponent $A_d(t)$ depends weakly on time and is completely unimportant
for $\tau_{\varphi}$ determined solely by the function $f_d(t)$ in the
exponent. However, the pre-exponent $A_d(t)$ {\it does} contribute to a short
time expansion of $\langle P(t)\rangle_{\rm diff}$ (equivalent to the
expansion in the interaction). In the first order of this expansion 
and at $T=0$ the leading term $\propto t$ from the exponent 
(\ref{fquantum}) will be exactly cancelled by the analogous term 
from $A_d(t)$. The remaining term $\propto \sqrt{t}\ln t$ from 
(\ref{fquantum}) which also produces (weaker) dephasing is not cancelled 
\cite{GZ99} but one can mistreat or even miss this
term completely if in addition one would expand in $1/Tt$ or use the
golden rule approximation. In the latter case one would obtain an
incorrect conclusion $1/\tau_{\varphi}=0$ at $T=0$.

Finally, let us mention that the ``$P(E)$-theory'' is known to well
describe the results of various experiments (see e.g. \cite{Clarke}).
Here we demonstrated that the low temperature saturation of the 
junction conductance 
\cite{Clarke} and of the time $\tau_{\varphi}$ extracted from the 
magnetoconductance measurements \cite{Webb} is of exactly the same 
physical nature. In both cases it is caused by the 
electron-electron interaction.

\end{document}